\documentclass[aps,pre,preprint]{revtex4}
\usepackage{epsfig}
\usepackage{graphicx}
\usepackage{amsmath,amssymb,amsfonts,latexsym}
\usepackage{graphicx}

\usepackage{epsfig,amssymb,amsfonts,verbatim}
\usepackage{amsmath}
\usepackage{latexsym}
\usepackage{amsfonts}
\usepackage{amssymb}
\usepackage{color}
\def\bfl{\begin{flushleft}}
\def\efl{\end{flushleft}}
\def\bfr{\begin{flushright}}
\def\efr{\end{flushright}}
\def\bc{\begin{center}}
\def\ec{\end{center}}

\def\ba{\begin{eqnarray}}
\def\ea{\end{eqnarray}}
\def\baa#1{\begin{array}{#1}}
\def\eaa{\end{array}}
\def\bw{\begin{widetext}}
\def\ew{\end{widetext}}

\def\text#1{\mbox{#1}}

\Large
\begin{document}

\title{Quantum thermodynamics at critical points during melting and solidification processes}
\author{A D Arulsamy$^*$}
\affiliation{F-02-08 Ketumbar Hill, Jalan Ketumbar, Taman Cheras, 56100 Kuala-Lumpur, Malaysia}



\maketitle

~\\ \textbf{Abstract:} We systematically explore and show the existence of finite-temperature continuous quantum phase transition ($^{\rm C}_T$QPT) at a critical point, namely, during solidification or melting such that the first-order thermal phase transition is a special case within $^{\rm C}_T$QPT. In fact, $^{\rm C}_T$QPT is related to chemical reaction where quantum fluctuation (due to wavefunction transformation) is caused by thermal energy and it can occur maximally for temperatures much higher than zero Kelvin. To extract the quantity related to $^{\rm C}_T$QPT, we use the ionization energy theory and the energy-level spacing renormalization group method to derive the energy-level spacing entropy, renormalized Bose-Einstein distribution and the time-dependent specific heat capacity. This work unambiguously shows that the quantum phase transition applies for any finite temperatures. ~
$^*$Corresponding Author, E-mail: sadwerdna@gmail.com
\newpage
\section*{1. Introduction}

A critical point here refers to a melting point of ice or solid such that at the critical point state one does not have a well defined phase of matter or phase boundary (both solid and liquid phases coexist with dynamic phase boundary between them). The fluctuation at a critical point during finite-temperature continuous quantum phase transition ($^{\rm C}_T$QPT) is specifically called quantum fluctuation (even for finite temperatures) because it is related to the properties of electrons at a constant temperature, and therefore, it is still subject to the Heisenberg uncertainty principle. For example, little did we know that physically and logically, not all quantum fluctuation ($\omega_{1,2,\cdots}$) has become `irrelevant' when $T > 0$K~\cite{sachBOOK,tvojta,mvojta,lava}. In other words, not all $\omega \rightarrow$ maximum when $T \rightarrow 0$K. If the statement---all quantum fluctuation become maximum only when $T \rightarrow 0$K is true, then chemical reaction cannot exist. The argument that a chemical reaction is a quantum mechanical effect can be proven if we can prove the existence of quantum fluctuation (due to wavefunction transformation) that becomes large for finite temperatures (or when $T > 0$K). The details on zero-temperature quantum phase transition ($_{\rm 0K}$QPT) can be found elsewhere~\cite{sachBOOK,tvojta}. The quantum fluctuation responsible for $^{\rm C}_T$QPT occurring at a finite temperature quantum critical point ($_T$QCP) is associated to quantitative changes to the wavefunction, which is also known as the wavefunction transformation~\cite{andQAT}.

Here, we systematically explore the notion of thermal phase transition beyond the structural phase transition~\cite{isik} and prove the existence of $^{\rm C}_T$QPT during melting and solidification (due to chemical reaction) and the standard first-order thermal phase transition (TPT) as a special case within $^{\rm C}_T$QPT. To achieve the above objectives, we first need to introduce and explain the ionization energy theory (IET), and this is followed by in-depth analysis on the first-order thermal phase transition in molten alkali halides, including why and how this transition is related to the finite temperature quantum phase transition. Subsequently, we dig deep into the details of quantum phase transition and quantum fluctuation with respect to IET, involving wavefunction transformation and electronic phase transition. Along the way, the theory is also discussed and crosschecked whenever necessary against the available experimental observations. Our primary results obtained from the proofs of points stated above are related to new physics beyond the traditional thermal phase transition and the zero-temperature quantum phase transition. 

\section*{2. Theoretical details}

\subsubsection*{\textit{2.1. Ionization energy theory}}

The ionization energy theory can be compactly captured by the IET-Schr${\rm \ddot{o}}$dinger equation~\cite{andPRA,andAOP},
\begin {eqnarray}
{\rm i}\hbar\frac{\partial \Psi(\textbf{\textit{r}},t)}{\partial t} &=& \bigg[-\frac{\hbar^2}{2m}\nabla^2 + V_{\rm IET}\bigg]\Psi(\textbf{\textit{r}},t) \nonumber \\&=& H_{\rm IET}\Psi(\textbf{\textit{r}},t) = (E_0 \pm \xi)\Psi(\textbf{\textit{r}},t), \label{eq:1}
\end {eqnarray}  
and the ionization energy approximation,
\begin {eqnarray}
\xi^{\rm quantum}_{\rm system} \propto \xi^{\rm constituent}_{\rm atom}. \label{eq:2}
\end {eqnarray}  
Here $\xi$ is the energy-level spacing or the ionization energy, $H_{\rm IET}$ and $V_{\rm IET}$ are the exact Hamiltonian and potential term, respectively, $\hbar$ is the Planck constant divided by 2$\pi$, $m$ denotes electron mass and $E_0$ is ground state energy for temperature ($T$) equals zero Kelvin, in absence of external disturbances such that $\xi$ determines the effects of any perturbation or disturbances and finally, the $\pm$ sign refers to electrons and holes, respectively. Add to that, only the true and real (not a guessed) wavefunction denoted by $\Psi(\textbf{\textit{r}},t)$ can properly capture the properties of electrons. The notation, $\xi^{\rm quantum}_{\rm system}$ denotes the real ionization energy of a quantum system, while $\xi^{\rm constituent}_{\rm atom}$ is the real atomic ionization energy. Accurate values for $\xi^{\rm constituent}_{\rm atom}$ can be directly obtained from any atomic spectra database, or routinely calculated from the density functional theory or other quantum chemical methods by using guessed wavefunctions and adjustable parameters.

Firstly, Eq.~(\ref{eq:1}) is not specific such that $V_{\rm IET}$ needs to be expanded and specified for a given system and after finding $V_{\rm IET}$, we still need to solve the IET-Schr${\rm \ddot{o}}$dinger equation variationally for $\xi$. The potential energy can be specified following Fig.~\ref{fig:1}(a) and (b) where we have assumed the sketched atoms ($X$ and $Y$) to be polarizable, neutral and identical with discrete energy levels. Each atom, $X$ and $Y$ has one equally-polarizable electron, namely, $e_{X}$ and $e_{Y}$, respectively. Here, the negatively charged electrons and positively charged nuclei are arranged in this manner, $+$ $-$ $+$ $-$ along the same axis. In contrast, $V^{\rm e-ion}_{\rm Coulomb}$ is the renormalized screened Coulomb potential between atomic $X$ and $Y$, which has been added by hand into Eq.~\ref{eq:4000} (the first term on the right-hand side). This particular term ($V^{\rm e-ion}_{\rm Coulomb}$) imposes the condition that both $X$ and $Y$ are also single electron atoms, but with a little twist such that atomic $X$ has the least polarizable electron ($e_{X}$), while atomic $Y$ has the easily polarizable electron ($e_{Y}$). We will also denote atomic $X$ as an anion (namely, Cl), whereas atomic $Y$ represents a cation, namely, Li. The charge carried by each nucleus, $X$ and $Y$ in both sketches, (a) and (b) in Fig.~\ref{fig:1} is $+e$. In particular, for a two-atom system depicted in Fig.~\ref{fig:1}(a) and (b), the relevant potential energy ($V_{\rm IET}$) is the Ramachandran interaction (stronger than the usual van der Waals type)~\cite{rama,ramc,ramc2,ramplot} 
\begin {eqnarray}
&&V_{\rm IET} = \tilde{V}'_{\rm Ramachandran}(\xi) = \tilde{V}^{\rm e-ion}_{\rm Coulomb} + \tilde{E} - \hbar\sqrt{\frac{\tilde{k}}{m}}, \label{eq:4000}
\end {eqnarray}  
where
\begin {eqnarray}
&&\tilde{E} - \hbar\sqrt{\frac{\tilde{k}}{m}} = \frac{1}{2}\hbar\omega_0\bigg(\frac{1}{\sqrt{2}} - 1\bigg)\bigg\{\exp{\bigg[\frac{1}{2}\lambda\xi_{X}\bigg]} + \exp{\bigg[\frac{1}{2}\lambda\xi_{Y}\bigg]}\bigg\}, \label{eq:4001}
\end {eqnarray}  
and
\begin {eqnarray}
\tilde{V}^{\rm e-ion}_{\rm Coulomb} &=& \frac{(-e)(+e)}{4\pi\epsilon_0|\textbf{\textit{R}}_{X} - \textbf{\textit{r}}_{Y}|}\bigg\{\exp{\big[-\mu r_{X}e^{-\frac{1}{2}\lambda\xi_{X}}\big]}\bigg\}. \label{eq:4002}
\end {eqnarray}  
From~\cite{andAOP}
\begin {eqnarray}
\tilde{k} = k\exp{[\lambda\xi]}, \label{eq:400k}
\end {eqnarray}  
where $\tilde{E}$, $\tilde{V}^{\rm e-ion}_{\rm Coulomb}$ and $\tilde{k}$ are the renormalized energy, electron-ion Coulomb potential and interaction potential constant, respectively, $\mu$ is the screening constant of proportionality, $\lambda = (12\pi\epsilon_0/e^2)a_{\rm B}$, $a_{\rm B}$ is Bohr radius of atomic hydrogen, $\textbf{\textit{R}}_{X}$ and $\textbf{\textit{r}}_{Y}$ are coordinates for nucleus $X$ and electron $e_{Y}$, respectively. Here, $\hbar(\tilde{k}/m)^{1/2} = \hbar\tilde{\omega}_0$, $\hbar\tilde{\omega}_0$ is the renormalized energy when $V_{\rm IET} = 0$, and $\tilde{V}'_{\rm Ramachandran}(\xi)$ is negative guaranteed by $(1/\sqrt{2}) - 1$ in Eq.~(\ref{eq:4001}) and $-e$ in Eq.~(\ref{eq:4002}). See Ref.~\cite{rama} for the derivation of Eq.~(\ref{eq:4000}). The term on the right-hand side of Eq.~(\ref{eq:4001}) assumes that both atoms, $X$ and $Y$ are identical, hence their electrons are equally polarizable. In Eq.~(\ref{eq:4002}) however, we have a unidirectional electron-ion attractive interaction between the easily polarizable $e_{Y}$ and nucleus $X$, screened by $e_{X}$ \textit{via} $\xi_{X}$ (see the term in the curly bracket in Eq.~(\ref{eq:4002})). The imposed asymmetric polarizability is captured by $\tilde{V}^{\rm e-ion}_{\rm Coulomb}$ such that the respective atomic $X$ and $Y$ represent an anion (least polarizable) and a cation (easily polarizable), which can be used to understand the transition from liquid ($V_{\rm IET}^{\rm liquid}$) to solid ($V_{\rm IET}^{\rm solid}$) phase. Here, the quantum phase transition originates from $V_{\rm IET}^{\rm liquid}(\xi_{\rm liquid}) \rightarrow V_{\rm IET}^{\rm solid}(\xi_{\rm solid})$, which requires wavefunction transformation because we need to transform $\xi_{\rm liquid} \rightarrow \xi_{\rm solid}$ accordingly.

To avoid calculations on the basis of variational principle that require guessed wavefunctions and variationally adjustable parameters, we have devised an alternative first principles approach where one just need to use the analytic ionization energy approximation (Eq.~(\ref{eq:2})) to evaluate the physical properties of a particular quantum system such that any changes to $\xi$ can be traced back to a renormalized physical parameter within the energy-level spacing renormalization group method~\cite{andAOP,shank1,shank2,shank3}. For example, the quantum fluctuation introduced above can be written in the form, $\hbar\omega = \xi$, and for different compositions ($y_1, y_2, \cdots$) the energies of fluctuation read, 
\begin {eqnarray}
&&\hbar\omega(y_1) < k_{\rm B}T_{\theta}, \label{eq:01} \\ && \hbar\omega(y_2) = k_{\rm B}T_{\theta}, \label{eq:02} \\ && \hbar\omega(y_3) > k_{\rm B}T_{\theta}, \label{eq:03}
\end {eqnarray}  
where $\hbar\omega(y) \neq k_{\rm B}T_{\theta}$ is the doping ($y$)-dependent energy that does not cause any fluctuation, and on the other hand, $\hbar\omega(y_2) = k_{\rm B}T_{\theta}$ is the energy at finite-temperature quantum critical point ($_T$QCP), where $\xi$ fluctuates due to transforming wavefunction~\cite{andQAT} at a critical point. Here, $y_1$, $y_2$ and $y_3$ are related to different chemical compositions, while $\theta$ represents a certain physical property under investigation. For example, for melting, $T_{\theta} = T^{\rm melting}_{\rm point}$. In contrast, for the usual $_{0\rm K}$QCP, the relevant inequality is simply defined~\cite{sachBOOK,tvojta} to be $\hbar\omega(y_1) < \hbar\omega(y_2) < \hbar\omega(y_3)$ because $ T_{\theta} = 0$K, or generally one writes $k_{\rm B}T < \hbar\omega$ for small temperatures not far from 0K to justify the validity of $_{0\rm K}$QCP. 

To understand how thermal energy initiates quantum fluctuation during melting and solidification processes, we have to evaluate the relationship between thermally driven first-order TPT and $^{\rm C}_T$QPT. Here, $^{\rm C}_T$QPT requires quantitative changes to the properties of electrons (due to changes in their energy levels) that need some qualitative and/or quantitative changes to their wavefunctions. These changes are driven by thermal energy, not by the temperature-independent external tuners such as pressure, electric and magnetic fields. In contrast, the commonly accepted quantum phase transition for $T = 0$K ($_{\rm 0K}$QPT) is driven by the temperature-independent external tuners, and requires quantitative changes to the properties of electrons (due to changes in their energy levels) that also need some qualitative and/or quantitative changes to their wavefunctions. The changes for $T = 0$K are driven by temperature-independent external tuners alone, not by thermal energy. But this does not imply that the thermal energy cannot be the cause for $^{\rm C}_T$QPT. Hence, we should not reserve all `quantum phase transitions' exclusively for transitions driven by temperature-independent external tuners (for $T = 0$K), and not for transitions driven by thermal energy for $T > 0$K. 

The above reservation is somewhat naive because external drivers can be anything, including thermal energy, for as long as these drivers can initiate quantum fluctuation for any $T \geq 0$K. Of course, a given QPT can be driven by thermal energy, with further assistance from other temperature-independent tuners, or a QPT can be suppressed if one external tuner competes with other tuners. In this case, it is just a matter of finding which tuner is in command of which interaction and quantum fluctuation.

We also note here that QPT is never proven to exist only for $T \approx 0$K. In fact, the existence of $_{\rm 0K}$QPT does not imply QPT cannot exist for $T > 0$K where the latter implication (QPT cannot exist for $T > 0$K) has become an implicit assumption nowadays. To understand this assumption, we refer to Ref.~\cite{snow}. Snow et al.~\cite{snow} have obtained the charge-density-wave (CDW) phase transition by measuring the low-temperature pressure-dependent Raman scattering~\cite{raman,raman2,raman3,raman4}. This CDW transition captures the transition from `crystalline' phase at low pressure to `disordered' phase for high pressure by means of changing Raman-scattering-induced optical and amplitude modes~\cite{snow}. Disordered phase here means that there is no long- or short-range CDW order. 

Anyway, the CDW transition temperature ($T_{\rm CDW}$) is found to be about 200K, but only the pressure-dependent CDW transition for $T = 3.5$K is recognized as a $_{\rm 0K}$QPT, which is an assumption restricted by an \textit{ad hoc} condition, $k_{\rm B}T < \hbar\omega$. We claim here that if the above pressure-dependent CDW transition, for any constant temperatures (including $T = 0$K), is accompanied by this continuous transformation, $\xi(P>P_{\rm CDW}) \rightarrow \xi(P<P_{\rm CDW})$ such that $\xi(P>P_{\rm CDW}) \neq \xi(P<P_{\rm CDW})$, then the CDW transition even for $T \approx 200$K can be recognized as a type of QPT. Hence, our primary aim here is to prove that thermal energy can cause quantum phase transition such that the condition $k_{\rm B}T < \hbar\omega$ is no longer a restriction to study QPT where the stated \textit{ad hoc} condition is just an assumption. We apply our theory to explain the doping- and temperature-dependent melting process in alkali halides and the freezing phenomenon in water in the presence of hydrogen bonds.

\subsubsection*{\textit{2.2. Thermal and quantum fluctuations}}

Here, we first justify the reason why we have selected salts made up of single-valent cations and anions from group 17 of the periodic table. Table~1 lists the diatomic bonding energies and melting points of these salts such that the melting points decrease systematically with decreasing bonding energies from F to I. Such a decreasing trend can be precisely overlapped with decreasing ionization energies of anions, namely, $\xi_{\rm F^+}$(1681 kJmol$^{-1}$) $>$ $\xi_{\rm Cl^+}$(1251 kJmol$^{-1}$) $>$ $\xi_{\rm Br^+}$(1140 kJmol$^{-1}$) $>$ $\xi_{\rm I^+}$(1008 kJmol$^{-1}$). The values in these inequalities are given in Table~2, follow the values marked with `` * ''. The ionization energy values were obtained from Ref.~\cite{web}. Interestingly however, for a given anion, the decreasing melting points do not agree with the ionization energy values for cations: Li, Na, K and Rb (follow their values listed in Table~2, marked with `` $\dag$ ''). 


For example, the melting points for NaF (996$^{\rm o}$C) $>$ KF (858$^{\rm o}$C) $>$ LiF (848$^{\rm o}$C) $>$ RbF (833$^{\rm o}$C) cannot be overlapped with $\xi_{\rm Li^+}$(520 kJmol$^{-1}$) $>$ $\xi_{\rm Na^+}$(496 kJmol$^{-1}$) $>$ $\xi_{\rm K^+}$(419 kJmol$^{-1}$) $>$ $\xi_{\rm Rb^+}$(403 kJmol$^{-1}$). This is not surprising because melting is a process directly proportional to bonding strengths (see Table~1), which have been discussed earlier~\cite{andCPC} within IET where large ionization energy values of cation-like ions ($\xi_{\rm C^{2+}} < \xi_{\rm O^{2+}}$) in a given molecule (C$^{2+}$O$^{2-}$ or O$^{2-}_{(1)}$O$^{2+}_{(2)}$) do not necessarily imply stronger bonds. Fortunately, we have proven that a stronger bond is predictable from IET if one considers the ionization energies for anions (say O in H$_2$O), because oxygen defines the ability to attract electrons from an atomic hydrogen~\cite{andPCCP}. In this case, smaller $\xi_{\rm O}$ means weaker O$-$H bond, which is consistent with the above-stated ionization-energy inequalities for anions (group 17 elements) and the melting points. Here, even though melting points can be captured by changing the interatomic distance, but this does not imply that the valence electrons at the melting point of molten alkali halide are static because these valence electrons are always dynamic such that they constantly rearrange themselves due to chemical reaction (during melting or solidification) in the presence of mobile ions or atoms.        


In Table~2, we have deliberately selected systems consisting of single valent cations and anions from group 17 to avoid effects from different electronic interactions due to different number (and type) of constituent atoms in a given molecule. To understand this point, we again use molecular systems, in which, for a NO$_2$ molecule, N acts as a cation, while O as an anion, which means that we need to consider the valence state, 4+ since four electrons have been transfered from N$^{4+}$ to O$^{2-}_{2}$. This electron-transfer is due to $\xi_{\rm N^{4+}} < \xi_{\rm O^{4+}}$ (see Table~1). In contrast, N is an anion in NH$_3$ molecule due to $\xi_{\rm N^{+}} > \xi_{\rm H^{+}}$ where three electrons from three hydrogen atoms are transfered to nitrogen, giving N$^{3-}$H$^{+}_3$. Thus far, the analysis is correct. However, the polarizability for molecule NO$_2$ \textit{does not} solely depend on these four electrons contributed by nitrogen if we compare N$^{4+}$O$^{2-}_2$ with N$^{3-}$H$^{+}_3$ because we cannot simplify the analysis by comparing $\xi_{\rm N^{4+}}$ with $\xi_{\rm H^{+}}$ (3 electrons contributed by three hydrogen ions) only. If we do so, then obviously we have $\xi_{\rm N^{4+}} > \xi_{\rm H^{+}}$ that falsely allows us to conclude $\alpha_d^{\rm NO_2} < \alpha_d^{\rm NH_{3}}$ because $\alpha_d \propto \exp({-\xi})$ where $\alpha_d$ is the displacement polarizability. 

In other words, we cannot use the above inequality ($\xi$) to directly compare $\alpha_d$ between NO$_{2}$ and NH$_{3}$ molecules because N acts as a cation in a NO$_{2}$ molecule, while it is an anion in NH$_{3}$. This implies that we need to take both the cationic- and anionic-effect into account explicitly for an accurate logical analysis, which have been correctly invoked in Ref.~\cite{andPCCP}. However, the anionic-effect can be neglected when we compare CO with O$_2$ because in these molecule, oxygen is the anion and therefore $\alpha_d^{\rm CO} < \alpha_d^{\rm O_{2}}$ is valid~\cite{andCPC}. On the other hand, the justification required to neglect the cationic-effect for doped-Pnictide superconductors is given in Ref.~\cite{andJSNM}. Finally, taking both cationic- and anionic-effect into account means that for large (many-electron) cations such as K ($Z$ = 19) and Rb ($Z$ = 37), there will always be some significant amount of polarization (due to large screening) from the outer core electrons, even though effectively, K$^+$ and Rb$^+$ are single-valent ions. This second- or third-electron polarization is negligible for few-electron atoms (due to small screening), namely, Li ($Z$ = 3) and interestingly, also for Na ($Z$ = 11). It should be clear now that in view of Eqs.~(\ref{eq:1}) and~(\ref{eq:2}), one cannot use electron affinities to evaluate melting or solidification process (see following sections).

Furthermore, we need smaller and larger ionization energies for cations and anions, respectively for stronger ionic bonds. Therefore, incorporating this additional amount of polarization (for cations) implies larger $\xi_{\rm cationic}$ or equivalently, smaller $\xi_{\rm anionic}$ that should reduce the melting point as has been observed experimentally (see Fig.~\ref{fig:2}(a)). In Fig.~\ref{fig:2}(a), melting points of Li(I, Br, Cl and F), Na(I, Br, Cl and F), K(I, Br, Cl and F) and Rb(I, Br, Cl and F) salts are plotted against the anionic (I, Br, Cl and F) ionization energies, while in Fig.~\ref{fig:2}(b), the vertical down-arrow (solidification) denotes the first-order thermal phase transition (from liquid to solid) for LiCl when temperature is reduced from $T$ $>$ 610$^{\rm o}$C to $T$ = 552$^{\rm o}$C. The horizontal arrow pointing left captures the continuous (purely electronic) quantum phase transition (from solid to liquid) at constant $T$ = 552$^{\rm o}$C by changing the chemical composition  systematically, LiCl$_{1-a_1}$Br$_{a_1-a_2}$I$_{a_2}$ such that LiCl $\rightarrow$ LiBr $\rightarrow$ LiI, $y_3 = 1-a_1$, $y_2 = a_1-a_2$ and $y_1 = a_2$. Importantly, the above correction does not apply to anions if a given system consists of mostly ionic bonds because anions are judged solely on their ability to attract electrons from cations. On the contrary, for systems with mostly covalent bonds, we need to consider both cationic- and anionic-effect simultaneously as carried out in Ref.~\cite{andPCCP}. The existence of this additional polarization is consistent with experimental results shown in Fig.~\ref{fig:2}(a) when one compares the slope, d$T_{\theta}$/d$\xi$ for salts containing Li and Na, for instance (d$T_{\theta}$/d$\xi$)$_{\rm Li}$ $\approx$ (d$T_{\theta}$/d$\xi$)$_{\rm Na}$, while (d$T_{\theta}$/d$\xi$)$_{\rm K}$ $\approx$ (d$T_{\theta}$/d$\xi$)$_{\rm Rb}$ (follow the solid lines in Fig.~\ref{fig:2}(a)). Here, we have shown that thermal energy can cause fluctuation in $\xi$, which could give rise to QPT. Later, we will expose why and how the fluctuation in $\xi$ (at a critical point) forces us to impose time-dependence into $\xi$.

\section*{3. Results and discussion}

\subsubsection*{\textit{3.1. Renormalized entropy and Bose-Einstein distribution}}

Figure~\ref{fig:2} shows direct proportionality between melting points and ionization energies of these constituent anions (I, Br, Cl and F). The respective vertical and horizontal arrows in Fig.~\ref{fig:2}(b) indicate first-order TPT and $^{\rm C}_T$QPT. In order to expose the existence of these phase transitions, we need to recall the first and second laws of thermodynamics. These two laws can be combined to obtain 
\begin {eqnarray}
{\rm d}U = \delta Q + \delta W = T{\rm d}S - P{\rm d}V, \label{eq:04}   
\end {eqnarray}    
in which, the change in the internal energy, d$U$ of a closed system equals the amount of heat ($Q$) absorbed and the amount of work ($W$) done by that system. Here, $\delta$ is not an exact differential because the changes in $Q$ and $W$ depend on the thermodynamic path (or independent of a particular system and process), and therefore, $Q$ and $W$ are not state functions. Here, $P$ and $V$ denote pressure and volume, respectively and we consider only reversible processes. Unlike $Q$ and $W$, the thermodynamic variables $U$, $P$ and $V$ are state functions that are unique for a given system and process. 

The second law is given by, $\delta Q = T{\rm d}S$ where $S$ is the entropy, another path-independent state function. All we need now is the relationship connecting $S$ to $\xi$ that can be obtained from the derivation of the ionization energy based Fermi-Dirac statistics ($i$FDS). Denoting $N$ as the total number of particles with $n_1$ particles have energy $(E_0 \pm \xi)_1$, $n_2$ particles with energy $(E_0 \pm \xi)_2$ and so on implies that $N = n_1 + n_2 + \cdots + n_m$. As a consequence, the number of ways for $q_1$ unoccupied quantum states to be arranged among $n_1$ particles is
\begin {eqnarray}
\Omega(n_1,q_1) = \frac{q_1!}{n_1!(q_1 - n_1)!}, \label{eq:05}   
\end {eqnarray}    
satisfying the condition that one empty quantum state can accommodate only one particle. Subsequently, the total number of ways for $q$ quantum states ($q = q_1 + \cdots + q_m)$ to be arranged among $N$ particles 
\begin {eqnarray}
\Omega(N,q) = \prod_i\frac{q_i!}{n_i!(q_i - n_i)!}. \label{eq:06}   
\end {eqnarray}    
The most probable configuration for certain $T$ can be obtained by maximizing the number of ways one can arrange $n_i$ particles in $q_i$ empty quantum states or, we need to maximize $\Omega(N,q)$ subject to the restrictive conditions, 
\begin {eqnarray}
\sum_in_i = N,~~~\sum_idn_i = 0, \label{eq:07}   
\end {eqnarray}    
\begin {eqnarray}
\sum_i(E_0 \pm \xi)_i = E_0 \pm \xi = E,~~~\sum_i(E_0 \pm \xi)_idn_i = 0 \label{eq:08}.   
\end {eqnarray}    
These conditions impose that the total energy, $E$ and the total number of particles, $N$ for a given system are always constant. Here, $E_0$ is the total energy for 0K. Using the standard procedure, and after some algebraic rearrangements, one gets
\begin {eqnarray}
\frac{N}{q} = \frac{1}{\exp{\big[\mu + \lambda_{\rm B}(E_0 \pm \xi)\big]} + 1}, \label{eq:11}   
\end {eqnarray}    
which is the $i$FDS. Taking $\exp{\big[\mu + \lambda_{\rm B}(E_0 \pm \xi)\big]} \gg 1$, $\mu = -\lambda_{\rm B} E^{0}_{\rm F}$ and dividing all the terms by $T$ will lead us to the energy-level spacing entropy
\begin {eqnarray}
S_{\xi} = -\frac{(E_0 \pm \xi) + E^{0}_{\rm F}}{T} = \frac{1}{(\lambda_{\rm B},\lambda) T}\ln{\frac{N}{q}}, \label{eq:12}   
\end {eqnarray}    
in which,  
\begin {eqnarray}
\lambda_{\rm B} = \frac{1}{k_{\rm B}T} ~~~{\rm for~constant}~~~ \xi ~~~{\rm and}  \label{eq:13}   
\end {eqnarray}    
\begin {eqnarray}
\lambda = \frac{12a_{\rm B}\pi\epsilon_0}{e^2} ~~~{\rm for~constant}~~~ T > 0. \label{eq:14}   
\end {eqnarray}    
The derivation for $\lambda_{\rm B}$ is well known in classical thermodynamics and is also given in Ref.~\cite{andPRA} within IET formalism, while the proof for $\lambda$ is available in Ref.~\cite{andPLA2}. Here, $E^{0}_{\rm F}$ denotes the Fermi level for $T = 0$K, independent of any disturbance. The entropy $S_{\xi}$ given in Eq.~(\ref{eq:12}) decreases logarithmically ($S_{\xi} \rightarrow -\infty$) when $N/q \rightarrow 0$ (ordered) such that, the inequality $0 < N/q < 1$ strictly corresponds to $-\infty_{\rm min} < S_{\xi} < 0_{\rm max}$. However, physically it is meaningless to have $-\infty_{\rm min}$ (unbounded from below) that will also lead to $(-S_{\xi})^{1/\texttt{m}}$ where $m$ is a natural number. Consequently, we will convert all our entropy equations to be positive definite as required in natural sciences. But in any case, within the set of real negative numbers, including zero ($S_{\xi} \in \mathbb{R}^-$) and Eq.~(\ref{eq:12}), we have the correct correspondence between $T \rightarrow 0$ and $S_{\xi} \rightarrow -\infty$ (ordered). Apart from that, it is straightforward to identify the intensive and extensive parameters in the above equations---for example, from Eq.~(\ref{eq:11}), one can readily verify $N$ and $q$ as extensive parameters, while $N/q$ and all the variables on the right-hand side of Eq.~(\ref{eq:11}) are intensive parameters. This means that Eq.~(\ref{eq:11}) captures the changes to the intensive parameters in the presence of external disturbances, namely, temperature and doping. 

The other relevant entropy is due to atomic-disorder, $S_{\texttt{D}} = k_{\rm B}\ln{\texttt{D}}$ where
\begin {eqnarray}
\texttt{D}(N_{\texttt{D}},q_{\texttt{D}}) = \prod_i\frac{(n_i + q_i - 1)!}{n_i!(q_i - 1)!} \approx \prod_i\frac{(n_i + q_i)!}{n_i!q_i!}, \label{eq:15}   
\end {eqnarray}
by taking $n_i \gg 1$, $q_i \gg 1$ and defining $N_{\texttt{D}}/q_{\texttt{D}}$ as the ratio of the total number of identical atoms and empty lattice sites. Here, $N_{\texttt{D}} = \sum_in_i$, $q_{\texttt{D}} = \sum_iq_i$, and $N_{\texttt{D}}/q_{\texttt{D}}$ corresponds to the probability of excited identical particles from their lattice sites. One of the two restrictive conditions needed to maximize Eq.~(\ref{eq:15}) remains the same (Eq.~(\ref{eq:07})) because the total number of particles (atoms) is also conserved. While the condition for the total energy has been renormalized, 
\begin {eqnarray}
\sum_i\big[Ee^{\frac{1}{2}\lambda\xi}\big]_in_i = \hbar\sqrt{\frac{\tilde{k}}{m}} = \tilde{E}, \label{eq:15x1}     
\end {eqnarray}    
to imply that the total energy (for the atoms, not electrons) is also conserved. Moreover, we have renormalized the total energy of identical atoms ($E$), not of the electrons, to obtain $\tilde{E}$ so that the electronic polarizability effect is taken into account. This means that the atoms are no longer considered to be the rigid vibrating ions, independent of their surrounding electrons. Somewhat similar to $i$FDS derived earlier, we also can derive the renormalized Bose-Einstein statistics ($r$BES)
\begin {eqnarray}
\frac{N_{\texttt{D}}}{q_{\texttt{D}}} = \frac{1}{\exp{\big[\lambda_{\rm B}Ee^{\frac{1}{2}\lambda\xi}\big]} - 1}. \label{eq:15x2}   
\end {eqnarray}    
The constant, $\alpha = \mu$ denotes the minimum energy prior to any excitation of atoms from their equilibrium lattice sites, somewhat similar to Fermi level for electrons, and obviously, it is zero here. It is important to note here that $q_{\texttt{D}} \geq N_{\texttt{D}}$ is required so that the probability is normalized to one, even if there can be any number of $n$ particles allowed to occupy a single empty quantum state, $q_i$. Similar to $S_{\xi}$, $S_{\texttt{D}} \in \mathbb{R}^-$. 

We are now ready to track the solidification of liquid LiCl from $T_2 > 610^{\rm o}$C to $T_1 = 552^{\rm o}$C as shown in Fig.~\ref{fig:2}(b). In this case, the dominant entropy change is due to the transition from the disordered (liquid state) to an ordered (solid) state in which, the liquid state is defined within 610$^{\rm o}$C $< T \leq T_2$ and the solid state is bounded within $T_1 \leq T <$ 610$^{\rm o}$C. The value 610$^{\rm o}$C ($T_{\theta}$) is the critical point, or the melting point of LiCl and $S_{\texttt{D}}^{\rm liquid} > S_{\texttt{D}}^{\rm solid}$ indicates the existence of TPT, qualitatively. To claim liquid has a larger entropy than solid means we have invoked an additional condition $|S_{\texttt{D}}^{\rm liquid} - S_{\texttt{D}}^{\rm solid}| > |S_{\xi}^{\rm liquid} - S_{\xi}^{\rm solid}|$ because the change of entropy due to electronic excitation is always smaller compared to an atomic-disorder induced entropy (due to broken translational symmetry) within a particular system. Of course, in the absence of this temperature-induced atomic-arrangement asymmetry, electron induced disorder or $S_{\xi}$ is the dominant entropy. 

Apparently, the change in entropy during the above-stated transition (solidification) is a first-order TPT because $S_{\texttt{D}:T > T_{\rm \theta}}^{\rm liquid}$ $>$ $S_{\texttt{D}:T < T_{\rm \theta}}^{\rm solid}$ such that there exist nonequilibrium entropy ($S_{\rm eqm}^{\rm non}$) that contribute to this inequality, which is classically undefined at the critical point, $T = T_{\rm \theta}$ = 610$^{\rm o}$C where both solid and liquid phases coexist. However, we can define $S_{\rm eqm}^{\rm non}$ with respect to time-dependent energy-level spacing, $\xi(t)$ and using Eq.~(\ref{eq:12}) at the critical point due to time-dependent processes of breaking and forming of bonds~\cite{andPCCP,andQAT} at the solid$|$liquid interface (we will revisit this issue in detail shortly) for $T = T_{\theta}$. In this case, $\xi$ continuously switches between $\xi^{\rm liquid}$ and $\xi^{\rm solid}$ for $T = T_{\theta}$. Consequently, one cannot write $S_{\texttt{D}}^{\rm solid} < S_{\rm eqm}^{\rm non} < S_{\texttt{D}}^{\rm liquid}$, which in turn implies that solidification and melting are first-order transitions. 

\subsubsection*{\textit{3.2. Renormalized specific heat}}

Traditionally, first-order TPT is defined to exist if there is quantitative discontinuity in the first derivative of any thermodynamic variable~\cite{gai}. For example, the first-derivative of heat capacity (a thermodynamic variable) is discontinuous at the critical point due to discontinuity in the entropy itself as stated above. Here, the constant-volume heat capacity and entropy relationship can be obtained from 
\begin {eqnarray}
{\rm d}S = \frac{{\rm d}U + {P\rm d}V}{T} = \frac{1}{T}\bigg[\frac{\partial U}{\partial T}\bigg|_V + 0 \bigg] = \frac{C_v}{T}{\rm d}T, \label{eq:16}
\end {eqnarray}    
using Eq.~(\ref{eq:04}) and the definition, $C_v = \partial U/\partial T|_V$, in which Eq.~(\ref{eq:16}) tells us nothing about the changes in $C_v$ during solidification or at the critical point, except that it diverges~\cite{crit}. In Eq.~(\ref{eq:16}), even though $T$ and $C_v(T)$ are intensive parameters, $S$ is an extensive parameter because $S$ is also influenced by other independent physical processes (not taken into account) where the change in entropy (d$S$) in Eq.~(\ref{eq:16}) is solely due to d$T$ and $C_v(T)$. Anyway, on the way to understand the origin of first-order TPT (given above), we have pointed out the existence of a time-dependent $\xi(t)$, which fluctuates between $\xi^{\rm liquid}$ and $\xi^{\rm solid}$ for $T = T_{\theta}$ in the presence of both solid and liquid phases. Now, the above fluctuation in $\xi$ can be associated to the existence of $^{\rm C}_T$QPT at the critical point during melting or solidification in such a way that if we continue extracting heat from the LiCl system, then one has the liquid to solid transformation due to $[\xi^{\rm liquid} \rightarrow \xi^{\rm solid}] > [\xi^{\rm liquid} \leftarrow \xi^{\rm solid}]$ that freezes LiCl completely. The formation of bonds (or releasing of heat) during solidification occur continuously with respect to time that involves complete modification of the time-dependent many-body wavefunction ($\Psi(t)$). Within IET formalism, we do not require to know the changes in $\Psi(t)$, instead, we have $\xi(t)$ as the fundamental interaction-strength functional, which changes continuously and originates from the chemical constituents of a system. In other words, the chemical composition at the solid$|$liquid interface changes with time, giving rise to this correspondence rule, $\Psi(t) \rightarrow \xi(t)$. For a given temperature ($T \neq T_{\theta}$), if the chemical composition is $t$-independent (for example, in the solid or liquid phase) but $\textbf{r}$-dependent, then $\Psi(\textbf{r}) \rightarrow \xi(\textbf{r})$ where $\textbf{r}$ is the electron coordinate. In our earlier work~\cite{andAOP}, we have renormalized $C_v$ that can be exploited here to rewrite Eq.~(\ref{eq:16}) for $T = T_{\theta}$    
\begin {eqnarray}
{\rm d}\tilde{S}^{\rm non}_{\rm eqm}(t) = \frac{\tilde{C}_v(t)^{\rm non}_{\rm eqm}}{T_{\theta}}{\rm d}t = \frac{C_v(T)}{T_{\theta}}\exp{\bigg[-\frac{3}{2}\lambda\xi(t)\bigg]}{\rm d}t, \label{eq:17}
\end {eqnarray}    
which reinforces the logic that any change to the interaction strength during solidification, even though $T$-independent at the critical point, never ceases to be $t$-dependent, giving rise to $S_{\rm liquid} \rightarrow S_{\rm solid}$ and $C^{\rm liquid}_{v} \rightarrow C^{\rm solid}_{v}$ transitions. These $t$-dependent transitions, $\tilde{S}^{\rm non}_{\rm eqm}(t)$ and $\tilde{C}_v(t)^{\rm non}_{\rm eqm}$ can be readily captured via $\xi(t)$ without any divergence. In addition to that, we now claim that the solid $\rightleftharpoons$ liquid transition for $T = T_{\theta}$ is associated to $^{\rm C}_T$QPT. An essential point to note here is that we did not identify $C_v(T)$ in Eq.~(\ref{eq:17}) as the nonequilibrium specific heat at the critical point because $C_v(T)$ reduces to $C^{\rm liquid}_v(T)$ during solidification, whereas, $C_v(T) \rightarrow C^{\rm solid}_v(T)$ during melting. We also remind the readers to take note here that unlike $S_{\xi}$ and $S_{\texttt{D}}$, $\{\tilde{S},S\} \in \mathbb{R}^+$ because $\{N/q,N_{\texttt{D}}/q_{\texttt{D}}\} \in (0,1]$. Physically (stated earlier), we should not be comfortable with $\{S_{\xi},S_{\texttt{D}}\} \in \mathbb{R}^-$, hence we define an alternative equation
\begin {eqnarray}
S_{\xi}(+) = k_{\rm B}\ln{\bigg\{\frac{q}{\exp{\big[\mu + \lambda(E_0 \pm \xi)\big]} + 1}\bigg\}} = k_{\rm B}\ln{N}, \label{eq:18}   
\end {eqnarray}    
which can be obtained from Eq.~(\ref{eq:11}) and recall here that $N$ is the number of excited electrons. Similarly, 
\begin {eqnarray}
S_{\texttt{D}}(+) = k_{\rm B}\ln{\bigg\{\frac{q_{\texttt{D}}}{\exp{\big[\lambda_{\rm B}Ee^{(1/2)\lambda\xi}\big] - 1}}\bigg\}} = k_{\rm B}\ln{N_\texttt{D}}, \label{eq:19}   
\end {eqnarray}    
in this case, $N_\texttt{D}$ is the number of ions being displaced (or excited) from their crystallographic sites. Apart from Eqs.~(\ref{eq:18}) and~(\ref{eq:19}), we can also enforce positivity by rewriting Eq.~(\ref{eq:11}) such that 
\begin {eqnarray}
S_{\xi}^+ = -k_{\rm B}\frac{N}{q}\ln{\frac{N}{q}} ~~~~{\rm and}~~~~ S_{\texttt{D}}^+ = -k_{\rm B}\frac{N_{\texttt{D}}}{q_{\texttt{D}}}\ln{\frac{N_{\texttt{D}}}{q_{\texttt{D}}}}, \label{eq:19x}   
\end {eqnarray}    
because if $N_1/q > N_2/q$ and $0 \leq N_{1,2}/q \leq 1$, then
\begin {eqnarray}
-\frac{N_1}{q}\ln{\frac{N_1}{q}} > -\frac{N_2}{q}\ln{\frac{N_2}{q}}, \label{eq:19xx}   
\end {eqnarray}    
which will guarantee $S_1 \geq 0$, $S_2 \geq 0$ and $S_1 > S_2$ because $N/q$ increases or decreases faster than $\ln{(N/q)}$. For example, taking $N = 1$, $q = x$, $f(x) = 1/x $ and $g(x) = \ln{(1/x)}$, one can write $h[f,g] = (1/x)\ln{(1/x)}$, subsequently it is straightforward to show d$^{\texttt{m}}f$/d$x^{\texttt{m}}$ $>$ d$^{\texttt{m}}g$/d$x^{\texttt{m}}$ is always true when $x \geq 1$ and $\texttt{m} \geq 1$ where $x \in \mathbb{R}^+$ and $\texttt{m} \in \mathbb{N^*}$. In addition, $S_1 > S_2$ is consistent with increasing entropy if $N/q$ gets larger, and consequently we will always have $\{S_{\xi}(+),S^+_{\xi},S_{\texttt{D}}(+),S^+_{\texttt{D}}\} \in \mathbb{R}^+$ where $\mathbb{R}^+$ and $\mathbb{N}^*$ are the set of real positive numbers and positive integers, including zero, respectively. In summary, from the above analyses, we can construct the following statement, \\
\textbf{Statement 1}: Any first-order TPT has got to go through $^{\rm C}_T$QPT at constant $T$, \\
which will be shown to be true in the subsequent section. 

\subsubsection*{\textit{3.3. Finite-temperature continuous quantum phase transition}}

Thus far, we have exposed the existence of $^{\rm C}_T$QPT during LiCl solidification at $T = T_{\theta}$ (see the vertical arrow in Fig.~\ref{fig:2})(b), which is commonly accepted as the first-order TPT without going into the details of $^{\rm C}_T$QPT. Proving \textbf{Statement 1} requires one to track $^{\rm C}_T$QPT or the changes in $\xi(t)$ during solidification for constant $T = T_{\theta}$. Alternatively, one may also prove \textbf{Statement 1} by tracking the horizontal arrow pointing left shown in Fig.~\ref{fig:2}(b) by systematically changing the chemical composition via substitutional doping of Cl with Br and followed by I such that LiCl $\rightarrow$ LiBr $\rightarrow$ LiI at constant $T$ = 552$^{\rm o}$C, which is the melting point for LiBr. We will first address $^{\rm C}_T$QPT occurring during solidification (for constant $T$ = 610$^{\rm o}$C). 

The above $^{\rm C}_T$QPT during solidification of LiCl system can be captured by isolating the system right at the critical point ($T_{\theta}$ = 610$^{\rm o}$C) as shown schematically in Fig.~\ref{fig:3}. This system has been isolated at the critical point when $T$ = $T_{\theta}$ = 610$^{\rm o}$C such that the solid phase coexists indefinitely within the liquid phase with a dynamic phase boundary between those two phases. The magnified sketch beneath the main diagram shows the temperature differences between the solid phase ($T_{\rm sol}$) and its surrounding liquid phase ($T_{\rm liq}$) such that $T_{\rm liq} > T_{\theta} > T_{\rm sol}$ and $T_{\rm liq} - T_{\rm sol} \ll T_{\theta}$. At the boundary, the system is in extreme nonequilibrium where Li or Cl from the liquid phase may react to form a rigid ionic bond, and release heat ($-Q$) as energy. Conversely, the same amount of heat is absorbed ($+Q$) by Li or Cl in the solid phase so as to break free from the solid phase in order to be part of the liquid phase. Here, the liquid phase Li and Cl are indicated with filled circles, while the same elements in the solid phase are drawn as circles, but we did not bother to completely fill the solid phase with circles. These processes are due to chemical reactions between the highly-polarized Li (small $\xi_{\rm Li}$) and the least-polarized Cl (large $\xi_{\rm Cl}$) giving rise to the chemical association between them, forming the solid phase, which releases energy as heat ($-Q$) into the liquid phase. This energy transfer increases the kinetic energy of the liquid-phase Li and Cl that will eventually collide onto the solid-particle surface with increasing frequency, and thus could transfer this energy (+$Q$) back into the solid phase to initiate melting of LiCl solid. These two thermal-assisted processes (due to $\pm Q$) are the causes for this thermal-assisted (or finite-temperature) $^{\rm C}$QPT. In a macroscopic point of view, this particular isolated system, containing both solid and liquid phases, is in equilibrium because the average rate of melting and solidification is the same, hence the solid-to-liquid and liquid-to-solid transitions ($^{\rm C}_T$QPT) are in balance. But we need to go deeper to track the physico-chemical processes at the solid$|$liquid interface, which can be done with IET. For instance, the energy level spacings in the liquid and solid phases are $\xi^{\rm LiCl}_{\rm liquid}$ and $\xi^{\rm LiCl}_{\rm solid}$, respectively, and since the valence electrons in the liquid phase are all in the excited states (thermally polarized) then this implies $\xi^{\rm LiCl}_{\rm liquid}$ $<$ $\xi^{\rm LiCl}_{\rm solid}$. This inequality is strictly valid because the excited energy-level spacings are always narrower due to weak electron-electron ($e$-$e$) interaction in the presence of weak electron-nucleus ($e$-$nuc$) attraction. Conversely, large energy level spacings are inevitable for the energy levels close to nucleus~\cite{andPRA}.

Substituting $\xi^{\rm LiCl}_{\rm liquid}$ $<$ $\xi^{\rm LiCl}_{\rm solid}$ into Eq.~(\ref{eq:19x}) leads to $_{\rm liq}S^+_{\xi}$ $>$ $_{\rm sol}S^+_{\xi}$ and $_{\rm liq}S_{\texttt{D}}^+$ $>$ $_{\rm sol}S_{\texttt{D}}^+$, which then allow one to correctly conclude $S_{\rm liquid} > S_{\rm solid}$ where $S_{\rm liquid}$ = $_{\rm liq}S^+_{\xi}$ + $_{\rm liq}S_{\texttt{D}}^+$ and $S_{\rm solid}$ = $_{\rm sol}S^+_{\xi}$ + $_{\rm sol}S_{\texttt{D}}^+$. However, as we have noted earlier, $S_{\rm solid} < S_{\rm eqm}^{\rm non} < S_{\rm liquid}$ is invalid because we need to take the nonequilibrium effect (at the solid$|$liquid interface) into account. We now know that this effect occurs maximally at the critical point (when $T = T_{\theta}$), and at the solid$|$liquid interface where $S_{\rm eqm}^{\rm non}$ = $S_{\rm liquid}$ + $S_{\rm solid}$ + $S^{\rm inter}_{\rm face}$, and therefore $S_{\rm solid}$ $<$ $S_{\rm liquid}$ $<$ $S_{\rm eqm}^{\rm non}$ that guarantees the existence of first-order TPT, while $S^{\rm inter}_{\rm face}$ on the other hand, ensures the existence of $^{\rm C}_T$QPT, as well as the coexistence of both solid and liquid phases (see Fig.~\ref{fig:3}). This means that if $S^{\rm inter}_{\rm face} = 0$, then $T \neq T_{\theta}$ and consequently we have $S_{\rm eqm}^{\rm non}$ $\rightarrow$ $S_{\rm liquid}$ for $T > T_{\theta}$ or $S_{\rm eqm}^{\rm non}$ $\rightarrow$ $S_{\rm solid}$ for $T < T_{\theta}$. Using Eq.~(\ref{eq:17}), the non-divergent and renormalized 
\begin {eqnarray}
&&\tilde{C}_v(T_{\theta},t)^{\rm non}_{\rm eqm} = C_v(T)\exp{\Bigg\{-\frac{3}{2}\lambda\bigg[\sum_jJ_j\xi_{\rm solid} + (1 - J_j)\xi_{\rm liquid}\bigg]\Bigg\}}, \label{eq:21}
\end {eqnarray}    
in which, we have defined $\xi(t) = \sum_jJ_j\xi_{\rm solid} + (1 - J_j)\xi_{\rm liquid}$ where $j = \{t_1, t_2, \cdots, t_{\texttt{n}}\}$, $J = \texttt{X}^{\rm LiCl}_{\rm solid}/\texttt{X}^{\rm LiCl}_{\rm total}$ and $J \in [0,1]$. Here, $\texttt{X}^{\rm LiCl}_{\rm solid}$ is the number of Li and Cl atoms in the solid phase only, and $\texttt{X}^{\rm LiCl}_{\rm total}$ is the total number of Li and Cl atoms in both liquid and solid phases (or in a given system). Moreover, $J$ does not necessarily increase with time, if the heat-exchange fluctuates ($\pm Q$), and the total time between $t_1$ and $t_{\texttt{n}}$ is the time taken for a complete solidification of liquid LiCl for constant $T = T_{\theta}$. Eq.~(\ref{eq:21}) strictly implies that the magnitude of $\tilde{C}_v(t)^{\rm non}_{\rm eqm}$ does not change with $T$ only, but also with respect to changes in the interaction-strength parameter, $\xi(t)$. In particular, $\tilde{C}_v(t)^{\rm non}_{\rm eqm}$ can change due to changes in $\xi(t)$ from $\xi_{\rm liquid}$ to $\xi_{\rm solid}$ during solidification of LiCl liquid. In this latter case, the heat-exchange as depicted in Fig.~\ref{fig:3} is solely used to change the interaction strength via $\xi_{\rm solid} \rightleftharpoons \xi_{\rm liquid}$. 

If one employs an unrenormalized specific-heat equation, then it is always divergent for $T = T_{\theta}$ because it is undefined at this critical point. On the other hand, Eq.~(\ref{eq:21}) is well-defined such that $\xi$ can be exploited at melting points, without any divergence. For example, $\xi_{\rm liquid}$ and $\xi_{\rm solid}$ are constants for $T > T_{\theta}$ and $T < T_{\theta}$, respectively, and any heat exchanges that may exist between a given system and its surrounding only decrease or increase the system's temperature. At the critical point however, the heat exchanges are used only to modify the system's physico-chemical properties, hence, the system's temperature remains constant. In other words, the renormalized specific-heat equation (Eq.~(\ref{eq:21})) reduces to 
\begin {eqnarray}
&&\tilde{C}^{\rm liquid}_v(T) = C^{\rm liquid}_v(T)\exp{\bigg[-\frac{3}{2}\lambda\xi_{\rm liquid}\bigg]}, \label{eq:22}
\end {eqnarray}    
for liquid LiCl ($T > T_{\theta}$), and for solid LiCl ($T < T_{\theta}$), one just need to switch the label `liquid' with `solid' in the equation above. However, neither of these equations can be applied when $T = T_{\theta}$. We need Eq.~(\ref{eq:21}) for $T = T_{\theta}$. This completes the proof for Statement 1. 

If we now follow the horizontal arrow given in Fig.~\ref{fig:2}(b), we can show that both transitions (for vertical and horizontal arrows) are thermal-assisted $^{\rm C}_T$QPT, one occurring during solidification (discussed above for the vertical arrow) and the other originates due to changing chemical composition (at a constant temperature). For example, the first $^{\rm C}_T$QPT at the critical point during solidification (constant $T = T_{\theta}$) is initiated by the heat-exchange between liquid and solid LiCl, where d$S = \delta Q/T_{\theta}$, and from Eq.~(\ref{eq:12}), 
\begin {eqnarray}
{\rm d}|S_{\xi}| = {\rm d}S = \frac{\delta Q}{T_{\theta}} = \bigg|-\frac{(E_0 \pm {\rm d}\xi) + E^{0}_{\rm F}}{T_{\theta}}\bigg|, ~~|S_{\xi}| \in \mathbb{R}^+. \label{eq:23} 
\end {eqnarray}    
Equation~(\ref{eq:23}) correctly implies: (a) increasing absorption of heat in a system increases the entropy of that system, or \textit{vice versa}, (b) any amount of change in $Q$ ($\delta Q$) or $\xi$ (d$\xi$) corresponds accordingly to a change in entropy, as it should be, however, (c) the entropy of a given system decreases if $\xi$ increases, as strictly required by Eq.~(\ref{eq:12}), (d) systems with large $\xi$ need large amount of heat to initiate changes such as melting, for example, from Table~2, we know $[\xi_{\rm Br} < \xi_{\rm Cl}] \rightarrow [\xi_{\rm LiBr} < \xi_{\rm LiCl}]$ and therefore $T^{\rm LiBr}_{\theta} < T^{\rm LiCl}_{\theta}$, and (e) large amount of heat is required to be removed or added to initiate large changes to $\xi$, for example $\xi^{\rm LiCl}_{\rm liquid} \rightleftharpoons \xi^{\rm LiCl}_{\rm solid}$. Here, $\xi^{\rm LiCl}_{\rm liquid}$ is a constant, while $Q$ is the amount of heat removed to initiate the change, $\xi^{\rm LiCl}_{\rm liquid} -Q \rightarrow \xi^{\rm LiCl}_{\rm solid}$. Therefore, (d) refers to entropy-change of a system prior to any phase transition, \textit{i.e.}, for constant $\xi$ and $T < T_{\theta}$ or $T > T_{\theta}$. In contrast, (e) reveals changes in the intrinsic physico-chemical properties of a system due to changing $\xi$ when $T = T_{\theta}$. This means that d$|S_{\xi}|$ is only valid at the critical point, or when $T = T_{\theta}$ because $\xi$ does not change with $T$, but it changes significantly when the physico-chemical properties of a given system change.   

In order to understand the existence of thermal-assisted $^{\rm C}$QPT due to doping (follow the horizontal arrow in Fig.~\ref{fig:2}(b)), we increase Br content, replacing Cl for constant $T$ to obtain a system defined by LiCl$_{1-a_1}$Br$_{a_1}$. The inequality, $\xi_{\rm Br} < \xi_{\rm Cl}$ implies $\xi$ decreases with increasing Br content where this doping is carried out for constant $T$ = 552$^{\rm o}$C. Now, the critical point can be obtained for $a_1 = 1$ and at this point, $T = T_{\theta}$ = 552$^{\rm o}$C, which is the melting point of LiBr. As a consequence, if $\xi^{\rm LiCl}_{\rm solid} \rightarrow \xi^{\rm LiBr}_{\rm solid}$ is achieved through doping at $T = T^{\rm LiBr}_{\theta}$, then $Q$ activates the melting process, such that $\xi^{\rm LiBr}_{\rm solid} + Q \rightarrow \xi^{\rm LiBr}_{\rm liquid}$. However, we point out that $Q = \xi^{\rm LiBr}_{\rm liquid} - \xi^{\rm LiBr}_{\rm solid}$ is false. Note here that the above stated doping-induced $^{\rm C}_T$QPT can occur for any constant temperature, and even for $T = 0$K.

In summary, the so-called activation energy in this chemical association should be equal to $Q$, which is required to complete the liquid-to-solid transition, or \textit{vice versa}. In other words, to complete $\xi^{\rm LiCl}_{\rm liquid} \rightarrow \xi^{\rm LiCl}_{\rm solid}$ or $\xi^{\rm LiBr}_{\rm solid} \rightarrow \xi^{\rm LiBr}_{\rm liquid}$ transformation, significant changes to $t$-dependent many-body wavefunction are necessary, for example, $\Psi(t)^{\rm LiCl}_{\rm liquid} \rightarrow \Psi(t)^{\rm LiCl}_{\rm solid}$. The existence of such a transformation in $\Psi(t)$ has been proven within a new quantum adiabatic theorem developed for chemical reactions~\cite{andQAT}, which can be used to understand why the wavefunction of unreacted species need to be combined linearly or written in a different form for the compounds formed after chemical reaction. Interestingly, M$\ddot{\rm u}$ller and Goddard~\cite{vbt} have also pointed out such a case must exist during chemical reaction. Thus far, we have exploited the formalism developed for IET such that the only \textit{a priori} information one required to know is the atomic energy-level spacings listed in Table~2.

The technical steps taken to prove the existence of $^{\rm C}_T$QPT required us to first construct \textbf{Statement 1}. But before constructing the statement, we have first explained why and how the first-order thermal phase transition during melting or solidification of alkali halides can be exactly recaptured with quantum phase transition. For example, in Further Analysis I, we have explained that thermal energy at a critical point (during melting or solidification) can cause fluctuation in $\xi$, which could give rise to $^{\rm C}_T$QPT. To prove the fluctuation in $\xi$, we derived the entropy-change due to electron (see Eq.~(\ref{eq:11})) and ion (see Eq.~(\ref{eq:15x2})) at the critical point, which are the causes for $\xi$ to fluctuate. Equations~(\ref{eq:11}) and~(\ref{eq:15x2}) show that (i) zero-entropy is not required for $^{\rm C}_T$QPT and (ii) \textbf{Statement 1} can be constructed because the changes to the electronic energy levels (with contributions from electrons and ions) are caused by thermal energy leading to the wavefunction or $\xi_{\rm solid} \rightarrow \xi_{\rm liquid}$ transformation.

Subsequently, we moved on to show that the inequality, $S_{\texttt{D}}^{\rm solid} < S_{\texttt{D}}^{\rm liquid} < S_{\rm eqm}^{\rm non}$ is responsible to label melting or solidification as the first-order TPT. By digging deeper into this solidification process, we have proved that $\xi$ continuously switches between $\xi^{\rm liquid}$ and $\xi^{\rm solid}$ due to wavefunction transformation for $T = T_{\theta}$, which unambiguously implies that the above process (melting or solidification) satisfies $^{\rm C}_T$QPT such that TPT (independent of wavefunction) is a special case. Next, we derived the renormalized specific heat formula in terms of entropy, which is used to expose that the fluctuation in $\xi$ (at a critical point) forces us to impose time-dependence into the energy level spacing such that $\xi \rightarrow \xi(t)$ (see Eqs.~(\ref{eq:17}) and~(\ref{eq:21})), and these equations properly prove the correctness of \textbf{Statement 1}. In addition, we have explained why and how $S_{\xi}$ (see Eq.~(\ref{eq:23})) can be used to further support that $\xi(t)_{\rm liquid} \rightarrow \xi(t)_{\rm solid}$ is equivalent to $\Psi(t)_{\rm liquid} \rightarrow \Psi(t)_{\rm solid}$ that further reinforces the correctness of \textbf{Statement 1}. Finally, our theory and \textbf{Statement 1} has been used to explain the physical processes at the critical point in these systems, namely, alkali halides and water.

\subsubsection*{\textit{3.4. Water freezing}}

We apply Eq.~(\ref{eq:21}) to water-to-ice thermal phase transition during freezing. Figure~\ref{fig:4} sketches the specific heat versus temperature curve for constant pressure ($C_p(T)$) for water (see the dashed line). The dashed line denotes the usual $C_p(T)$ curve with a sudden drop ($C^{\rm water}_p(T) > C^{\rm ice}_p(T)$) at the freezing point. Whereas, the solid line in Fig.~\ref{fig:4} captures the whole mechanism of phase transition from water to ice, including the freezing curve occurring at the freezing point (273.16K). Obviously, the freezing curve is always hidden at the freezing point and appears as a sudden drop in $C_p(T)$ measurement, when temperature is lowered. Unfortunately, we do not know the explicit microscopic equation for $C_p(T) = {\rm d}\texttt{H}/{\rm d}T|_{\rm P}$ where $\texttt{H}$ is known as the enthalpy. However, the fundamental energy-level spacing renormalization factor (the exponential term in Eq.~(\ref{eq:21})) for $\tilde{C}_v(T_{\theta},t)^{\rm non}_{\rm eqm}$ should remain intact for $\tilde{C}_p(T_{\theta},t)^{\rm non}_{\rm eqm}$ such that     
\begin {eqnarray}
&&\tilde{C}_p(T_{\theta},t)^{\rm non}_{\rm eqm} = C_p(T)\exp{\Bigg\{-\frac{\eta}{2}\lambda\bigg[\sum_jJ_j\xi_{\rm ice} + (1 - J_j)\xi_{\rm water}\bigg]\Bigg\}}. \label{eq:27}
\end {eqnarray}    
The only change in the renormalizing factor is the numerical constant, 3 in $3\lambda/2$ (see the exponential term in Eq.~(\ref{eq:21})). The constant 3 here is unknown and therefore, we replace it with $\eta$ in Eq.~(\ref{eq:27}). This numerical constant, $\eta$ can only be obtained by solving $C_p(T) = {\rm d}\texttt{H}/{\rm d}T|_{\rm P}$, but it is irrelevant here. What we actually need are the renormalized specific heat equations for water and ice, 
\begin {eqnarray}
&&\tilde{C}^{\rm water}_p(T) = C^{\rm water}_p(T)\exp{\bigg[-\frac{\eta}{2}\lambda\xi_{\rm water}\bigg]}, \label{eq:28}
\end {eqnarray}    
\begin {eqnarray}
&&\tilde{C}^{\rm ice}_p(T) = C^{\rm ice}_p(T)\exp{\bigg[-\frac{\eta}{2}\lambda\xi_{\rm ice}\bigg]}. \label{eq:29}
\end {eqnarray}    

In Eqs.~(\ref{eq:28}) and~(\ref{eq:29}), only $C^{\rm water}_p(T)$ and $C^{\rm ice}_p(T)$ are $T$-dependent variables, and the rest are just $T$-independent variables or constants. This means that Eq.~(\ref{eq:28}) captures the specific heat for water, whereas, Eq.~(\ref{eq:29}) is for ice (see dashed and solid lines for water and ice in Fig.~\ref{fig:4}). The sketched freezing curve however, follows Eq.~(\ref{eq:27}) and is smooth here, implying there is no fluctuation in $Q$ during freezing, \textit{i.e.}, there is a continuous extraction of heat from the water-ice system. Recall here that the freezing curve in Fig.~\ref{fig:4} is not observable from the $C_p$ versus $T$ measurement alone, but exists as a sudden drop in the specific heat for $T = 273.16$K (freezing point) because $C^{\rm water}_p(T) > C^{\rm ice}_p(T)$ or $\xi^{\rm water} < \xi^{\rm ice}$. This is the reason why temperature is not a proper variable to monitor at the freezing point of any thermally-driven finite-temperature continuous quantum phase transition.

In summary, we have proven that $\xi(t)$ is the proper quantum variable to capture what is really happening at the freezing point. For example, at the freezing point, the heat that is being extracted from the water-ice system, activates the energy-level spacing transformation, $\xi^{\rm water} \rightarrow \xi^{\rm ice}$ or the wavefunction transformation $\Psi^{\rm water} \rightarrow \Psi^{\rm ice}$ due to the formation of permanent hydrogen bonds. This transformation results in $C^{\rm water}_p(T) > C^{\rm ice}_p(T)$ that implies $\xi^{\rm water} < \xi^{\rm ice}$. The latter inequality has its origin in hydrogen bonds. For example, both $\xi^{\rm water}$ and $\xi^{\rm ice}$ refer to energy-level spacings due to hydrogen bonds and not due to covalent bonds between O and H in a H$_2$O molecule. Obviously, the energy-level spacings due to hydrogen bonds are larger for ice compared to water because the hydrogen bonds in ice is permanently bonded (static) and therefore, its strength is larger in ice than in water. The hydrogen bonds in water phase is dynamic such that the bonds are broken and formed randomly. As such, indeed $\xi^{\rm water} < \xi^{\rm ice}$. Now for water vapor, all H$_2$O molecules are isolated, and they do not form any hydrogen bond. As a result, $\xi^{\rm vapor}$ refers to energy-level spacings due to covalent bonds (between O and H) within a H$_2$O molecule. Since covalent bonds are much stronger than hydrogen bonds, one has $\xi^{\rm water} < \xi^{\rm ice} < \xi^{\rm vapor}$ that correctly corroborates with experimental observations~\cite{tch,lish}, $C^{\rm water}_p$ (4.187 kJkg$^{-1}$K$^{-1}$) $>$ $C^{\rm ice}_p$ (2.108 kJkg$^{-1}$K$^{-1}$) $>$ $C^{\rm vapor}_p$ (1.996 kJkg$^{-1}$K$^{-1}$). This concludes our analytical proofs and technical analyses on H$_2$O system, again supported by experimental results.   

\section*{4. Conclusions} 
    
The finite-temperature continuous quantum phase transition ($^{\rm C}_T$QPT) has been formally proven to exist and to be responsible for the thermal phase transition during melting or solidification, as well as during substitutional doping for constant temperature. Along the way, we also have proven that first-order phase transition obtained from constant-pressure specific heat data between water and ice unambiguously satisfies the notion of $^{\rm C}_T$QPT developed here. In view of the analysis for alkali halide salts and water, we found that standard first-order thermal phase transition is a special case within $^{\rm C}_T$QPT. The equations developed herein allow one to obtain the precise physico-chemical mechanisms (without any divergences) right at the melting point of a given solid, beyond the usual physics of discontinuous (sudden change) thermal phase transition, specifically, in the first-derivative of specific heat capacity. Therefore, we can also claim that chemical reaction is a quantum critical point phenomenon that can be associated to the finite temperature continuous quantum phase transition.

\section*{Acknowledgments}

I thank Madam Sebastiammal Savarimuthu, Mr Arulsamy Innasimuthu, Madam Amelia Das Anthony, Mr Malcolm Anandraj and Mr Kingston Kisshenraj for their financial support and kind hospitality between August 2011 and August 2013. I also thank Dr Naresh Kumar Mani for his kind hospitality during my short stay in Cachan, France (March/April 2011) where part of this work was completed.

\newpage
~\\ \textbf{Table~1:} Experimental values for the melting points and diatomic bonding energies of salts obtained from Ref.~\cite{web}. The systematic decrease of melting points and diatomic bonding energies with respect to anions (from F to I with increasing $Z$) satisfy the decreasing ionization energies for the same anions, from F to I. See text and Table~2 (follow the values marked with `` * '') for details. \\ \\
\textbf{Table~2:} Averaged atomic ionization energies ($\xi$) for individual ions and their respective valence states ordered with increasing atomic number $Z$. All experimental ionization energy values were obtained from Ref.~\cite{web}. \\ \\
\textbf{Figure 1:} (a): The sketched atoms ($X$ and $Y$) are equally polarizable, neutral and identical with discrete energy levels. (b): Atomic $X$ has the least polarizable electron ($e_{X}$), while $e_{Y}$ from atomic $Y$ is easily polarizable. \\ \\
\textbf{Figure 2:} (a): Melting points of salts are plotted against the ionization energies of anions (I, Br, Cl and F). (b): Melting points of Li(I, Br, Cl and F) versus anions ionization energies. The vertical down-arrow (solidification) denotes the first-order thermal phase transition (TPT: from liquid to solid) for LiCl, while the horizontal left-arrow denotes quantum phase transition (QPT: from solid to liquid) at constant $T$ = 552$^{\rm o}$C. \\ \\
\textbf{Figure 3:} Both atomic Li and Cl in liquid phase are denoted with filled circles, while circles represent ionic Li and Cl in solid phase. \\ \\
\textbf{Figure 4:} A dashed line is sketched to represent the changes of constant-pressure specific heat ($C_p(T)$) with temperature for both water and ice. The solid line captures the same specific heat based on Eqs.~(\ref{eq:27}),~(\ref{eq:28}) and~(\ref{eq:29}), which gives a complete picture. 

\newpage
Table 1
\begin{table}[ht]
\begin{tabular}{l c c } 
\hline\hline 
\multicolumn{1}{l}{Salts}           &   Melting points 		& Diatomic bonding        \\  
\multicolumn{1}{l}{}                &   ($^{\rm o}$C)       & energies (kJmol$^{-1}$) \\  
\hline 

LiF                                   &  848   						&	577 					 \\  
LiCl                                  &  610	   	  		  & 469	   	  		 \\ 
LiBr																	&  552              & 419            \\
LiI																		&  469              & 345            \\\hline
NaF      															&  996							& 519						 \\
NaCl     															&  801							& 412						 \\	
NaBr		 															&  747							& 367						 \\
NaI			 															&  660							& 304            \\\hline
KF      															&  858							& 498						 \\
KCl     															&  771							& 433						 \\	
KBr		 															  &  734							& 380						 \\
KI			 															&  681							& 325            \\\hline
RbF      															&  833  					  & 494            \\ 
RbCl     															&  715	   	  		  & 428            \\ 
RbBr		 															&  682              & 381            \\
RbI			 															&  642              & 319            \\
\hline  
\end{tabular}
\label{Table:1} 
\end{table}

\newpage
Table 2
\begin{table}[ht]
\begin{tabular}{l c c c} 
\hline\hline 
\multicolumn{1}{l}{Elements}        &    Atomic numbers &  Valence     & $\xi$   \\  
\multicolumn{1}{l}{}                &   $Z$             &  states      & (kJmol$^{-1}$)\\  
\hline 

H                                   &  1   					    &  1+      & 1312 \\ 
Li                                  &  3	   	  			  &  1+      & 520$\dag$  \\ 
N																		&  7                &  1+      & 1402 \\
N																		&  7                &  4+      & 4078 \\
O                                   &  8	   	  			  &  1+      & 1314 \\ 
O                                   &  8	   	  			  &  2+      & 2351 \\ 
O                                   &  8	   	  			  &  4+      & 4368 \\ 
F                                   &  9	   	  			  &  1+      & 1681* \\ 
Na                                  &  11	   	  			  &  1+      & 496$\dag$  \\ 
Cl                                  &  17	   	  			  &  1+      & 1251* \\ 
K                                   &  19	   	  			  &  1+      & 419$\dag$  \\
Br                                  &  35	   	  			  &  1+      & 1140* \\ 
Rb                                  &  37	   	  			  &  1+      & 403$\dag$  \\ 
Sr                                  &  38	   	  			  &  2+      & 807 \\
I                                   &  53	   	  			  &  1+      & 1008* \\
La                                  &  57	   	  			  &  3+      & 1152 \\
\hline  
\end{tabular}
\label{Table:2} 
\end{table}

\newpage
\begin{figure}
\begin{center}
\scalebox{0.7}{\includegraphics{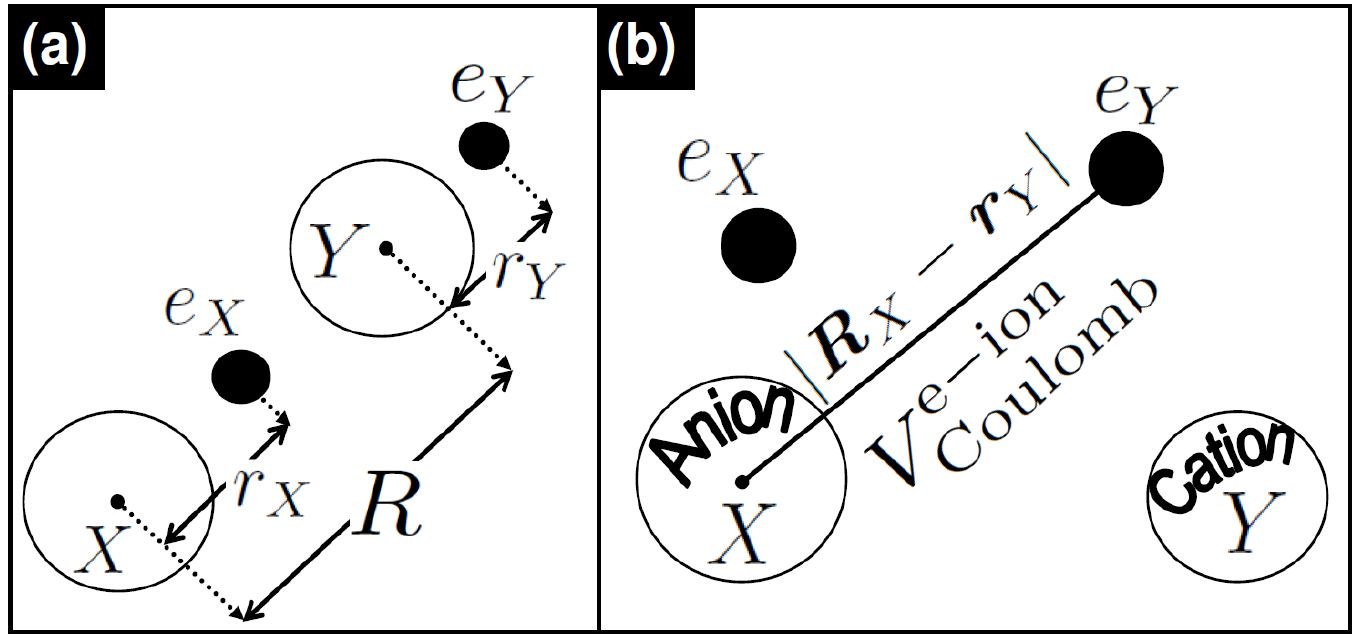}}
\caption{}
\label{fig:1}
\end{center}
\end{figure}

\newpage
\begin{figure}
\begin{center}
\scalebox{0.8}{\includegraphics{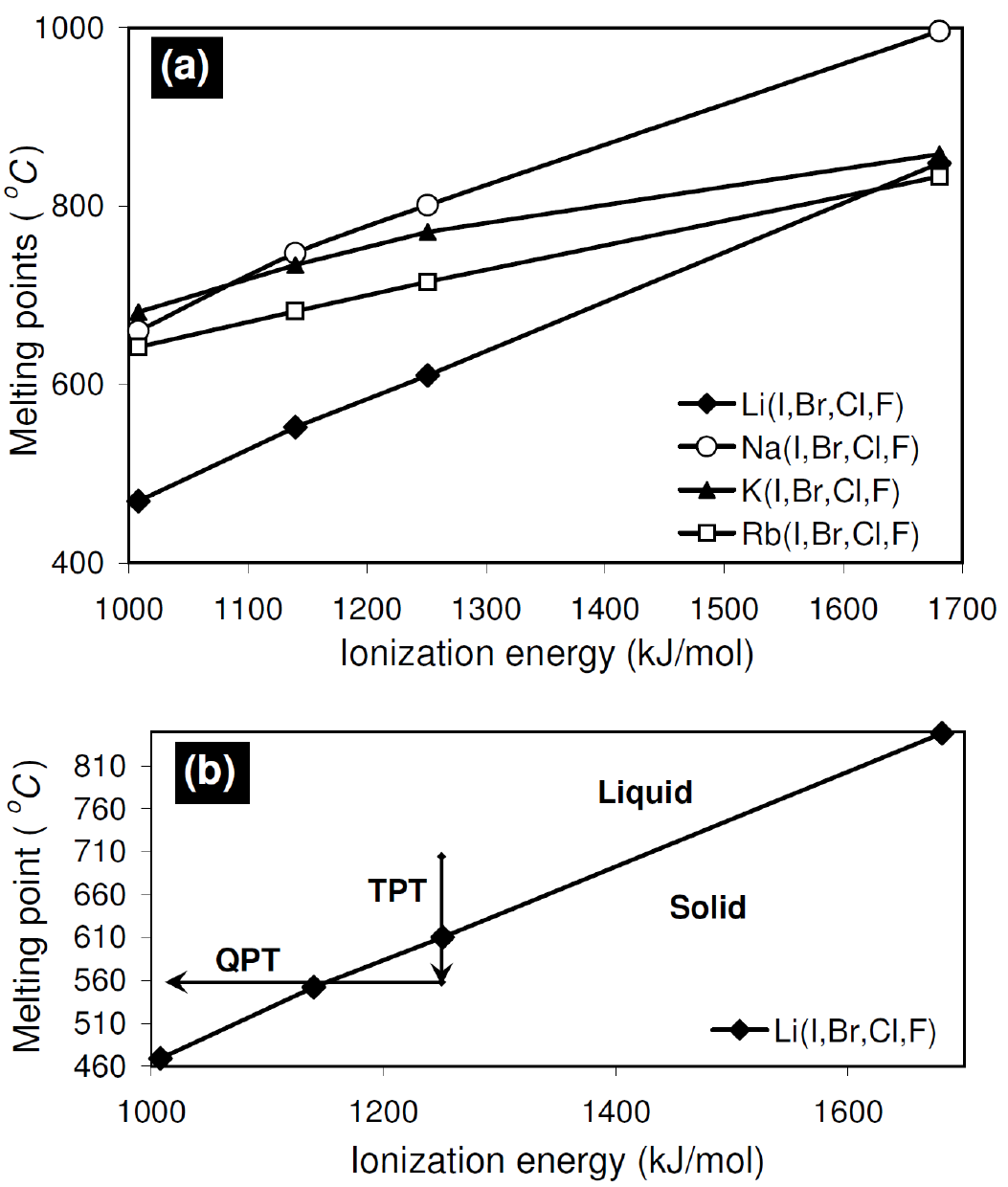}}
\caption{}
\label{fig:2}
\end{center}
\end{figure}

\newpage
\begin{figure}
\begin{center}
\scalebox{1.2}{\includegraphics{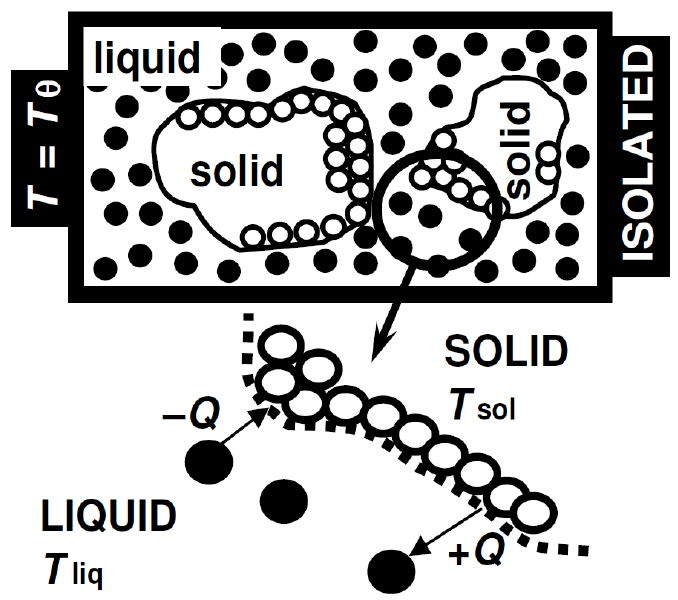}}
\caption{}
\label{fig:3}
\end{center}
\end{figure}

\newpage
\begin{figure}
\begin{center}
\scalebox{1.0}{\includegraphics{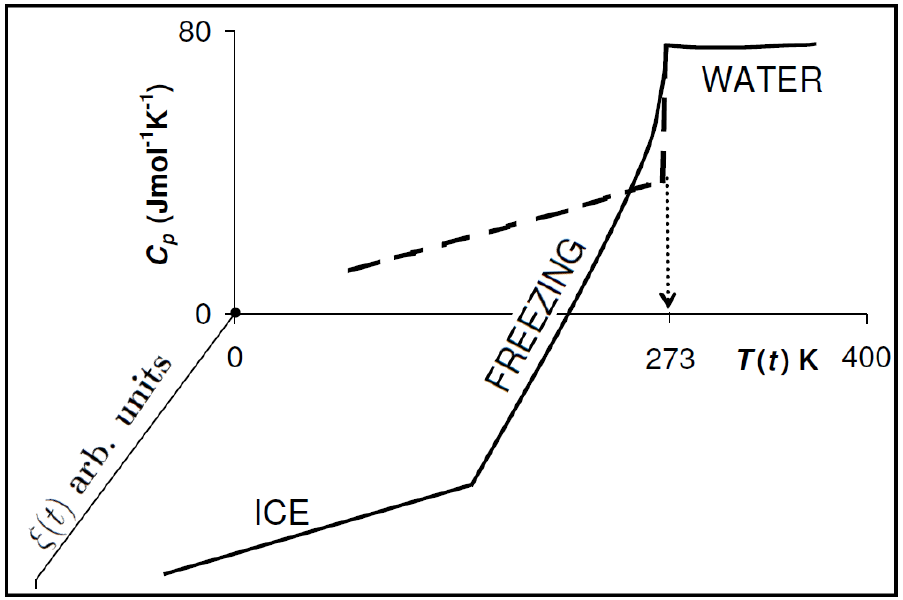}}
\caption{}
\label{fig:4}
\end{center}
\end{figure}


\begin{thebibliography}{plain}

\bibitem{sachBOOK} S Sachdev \textit{Quantum Phase Transitions} (New York : Cambridge University Press) Ch 1, p 3 (1999)

\bibitem{tvojta} T Vojta \textit{Ann. der Phys.} \textbf{509} 403 (2000)

\bibitem{mvojta} M Vojta \textit{Phil. Mag.} \textbf{86} 1807 (2006)

\bibitem{lava} M Lavagna \textit{Phil. Mag. B} \textbf{81} 1469 (2001)

\bibitem{andQAT} A D Arulsamy \textit{Prog. Theor. Phys.} \textbf{126} 577 (2011)

\bibitem{isik} F Isik, M A Sabaner, A T Akan and A Bayri \textit{Indian J. Phys.} \textbf{87} 241 (2013)

\bibitem{andPRA} A D Arulsamy \textit{Pramana J. Phys.} \textbf{74} 615 (2010)

\bibitem{andAOP} A D Arulsamy \textit{Ann. Phys.} \textbf{326} 541 (2011)

\bibitem{rama} A D Arulsamy \textit{J. Chem. Sci.} to be published (2014)

\bibitem{ramc} G N Ramachandran, V Sasisekharan and C Ramakrishnan \textit{J. Mol. Biol.} \textbf{7} 95 (1963)

\bibitem{ramc2} G N Ramachandran and V Sasisekharan \textit{Adv. Prot. Chem.} \textbf{23} 283 (1968)

\bibitem{ramplot} C Ramakrishnan and G N Ramachandran \textit{Biophys. J.} \textbf{5} 909 (1965)

\bibitem{shank1} R Shankar \textit{Physica A} \textbf{177} 530 (1991)

\bibitem{shank2} R Shankar \textit{Rev. Mod. Phys.} \textbf{66} 129 (1994)

\bibitem{shank3} R Shankar \textit{Phil. Trans. R. Soc. A} \textbf{369} 2612 (2011)

\bibitem{snow} C S Snow, J F Karpus, S L Cooper, T E Kidd and T C Chiang \textit{Phys. Rev. Lett.} \textbf{91} 136402 (2003)

\bibitem{raman} C V Raman \textit{Indian J. Phys.} \textbf{2} 395 (1928)

\bibitem{raman2} C V Raman \textit{Indian J. Phys.} \textbf{2} 387 (1928)

\bibitem{raman3} C V Raman and K S Krishnan \textit{Nature} \textbf{121} 501 (1928)

\bibitem{raman4} C V Raman and K S Krishnan \textit{Indian J. Phys.} \textbf{2} 399 (1928)

\bibitem{web} M J Winter $\langle$www.webelements.com$\rangle$ (2011)

\bibitem{andCPC} A D Arulsamy, K Eler$\check{\rm s}$i$\check{\rm c}$, M Modic, U Cvelbar and M Mozeti$\check{\rm c}$ \textit{Chem. Phys. Chem.} \textbf{11} 3704 (2010)

\bibitem{andPCCP} A D Arulsamy, Z Kregar, K Eler$\check{\rm s}$i$\check{\rm c}$, M Modic and U S Subramani \textit{Phys. Chem. Chem. Phys.} \textbf{13} 15175 (2011)

\bibitem{andJSNM} A D Arulsamy and K Ostrikov \textit{J. Supercond. Nov. Magn.} \textbf{22} 785 (2009)

\bibitem{andPLA2} A D Arulsamy \textit{Phys. Lett. A} \textbf{334} 413 (2005)

\bibitem{gai} S C Gairola \textit{Indian J. Phys.} \textbf{86} 967 (2012)

\bibitem{crit} L Cemi$\check{\rm c}$ \textit{Thermodynamics in Mineral Sciences} (Berlin : Springer) Ch 6, p 231 (2005)

\bibitem{vbt} R P M$\ddot{\rm u}$ller and W A Goddard \textit{Valence Bond Theory} (New York : Academic Press) Reprinted from the Encyclopedia of Physical Science and Technology (2002)

\bibitem{tch} V Tchijov \textit{J. Phys. Chem. Solids} \textbf{65} 851 (2004)

\bibitem{lish} S V Lishchuk, N P Malomuzh and P V Makhlaichuk \textit{Phys. Lett. A} \textbf{375} 2656 (2011)

\end{thebibliography}
\end{document}